\documentclass[12pt,english]{article}
\usepackage[T1]{fontenc}
\usepackage[latin9]{inputenc}
\usepackage{color}
\usepackage{amsmath}
\usepackage{amssymb}
\usepackage{stmaryrd}
\usepackage{geometry}
\usepackage{wasysym}
\usepackage{esint}

\makeatletter
\usepackage{braket}

\makeatother

\usepackage{babel}
\begin{document}
\title{A boundary term for open string field theory }
\author{Georg Stettinger}
\maketitle
\begin{center}
\textcolor{black}{{CEICO, Institute of Physics of the Czech Academy
of Sciences, }\\
\textcolor{black}{{} Na Slovance 2, 182 00 Prague 8, Czech Republic} }
\par\end{center}
\begin{abstract}
We consider Witten's open string field theory in the presence of a
non-trivial boundary of spacetime. For the kinetic term, we derive
a Gibbons-Hawking-type contribution that has to be added to the action
to guarantee a well-defined variational principle. The derivation
is done first in a heuristic way and then confirmed by a path integral
based approach using the CFT operator formalism. In the last section
we discuss the boundary contributions coming from the cubic vertex,
although it is problematic to apply consistent boundary conditions
on the string field due to the non-locality of the vertex.
\end{abstract}

\section{Introduction and motivation }

There exist many field theories defined on some spacetime manifold
$M$ whose precise definition becomes inconsistent if $M$ has a non-vanishing
boundary. It is often perfectly fine to ignore this issue because
the boundary is significantly far away and one is only concerned about
what happens in the bulk. However, in some situations the boundary
behaviour has to be taken account, not only for mathematical consistency
but also for physical reasons. The main example is Einstein gravity,
where the Einstein-Hilbert action 
\begin{equation}
S=\frac{1}{16\pi G_{N}}\int_{M}d^{4}x\,\sqrt{-g}R
\end{equation}
has to be supplemented with the Gibbons-Hawking term 
\begin{equation}
S_{GH}=\frac{1}{8\pi G_{N}}\int_{\partial M}d^{3}x\,\sqrt{\mid h\mid}K
\end{equation}
to give rise to a well-defined variational principle. Moreover, the
boundary term is necessary to produce the correct values of the ADM-mass
and black hole entropy \cite{Hawking1996}. Since string field theory
should contain gravity as one of its subsectors, it is natural to
assume that also string field theory should contain a non-trivial
boundary term. Additional evidence for why the boundary terms are
needed comes from the discrepancy between the results of \cite{Erler2022}
where it is shown that the on-shell action of closed string field
theory vanishes up to possible boundary contributions and \cite{Scheinpflug2023,Mazel2024},
from which it follows that the action for at least some classical
solutions corresponding to nearly marginal deformations in a linear-dilaton
background is non-zero.

Although both arguments concern closed string field theory, for simplicity
we want to start with open string field theory and analyze its variational
principle in the presence of a boundary in spacetime. The Witten action
\cite{Witten1986} of open string field theory\footnote{For a recent comprehensive review of the subject see \cite{Sen2024}.}
consists of two terms and reads 
\begin{equation}
S\left(\Psi\right)=-\frac{1}{2}\Braket{\Psi,Q\Psi}-\frac{1}{3}\Braket{\Psi,\Psi\ast\Psi}
\end{equation}
and we expect to get non-vanishing boundary contributions from both
of the terms. While for the kinetic term it will be quite straightforward
to obtain them, the cubic term is much more subtle due to its non-local
nature. The paper is organized as follows: After reviewing some general
facts on surface terms in field theory we will start by computing
the kinetic boundary term in a heuristic way using partial integration.
In section \ref{sec:Breaking-of-conformal} we will reproduce the
result in the operator formalism using the fact that the boundary
violates conformal invariance. Section \ref{sec:Cubic-boundary-term}
is of a more speculative character and deals with the cubic boundary
term: It is not clear if it even makes sense to have a localized boundary
where boundary conditions need to be applied in a non-local theory
(see \cite{Erbin2022,Moeller2002} for an analysis of the initial
value problem in non-local theories). We will discuss where boundary
terms are expected to appear and why their calculation is non-trivial. 

\section{General philosophy\label{sec:General-philosophy}}

Consider an ordinary field theory with a set of fields $\phi^{i}\left(x^{\mu}\right)$
and an action 
\begin{equation}
S\left[\phi\right]=\int_{M}d^{D}x\,\sqrt{-g}\,\mathcal{L}\left(\left[\phi\right],x^{\mu}\right),
\end{equation}
where $\left[\phi\right]$ denotes a dependence on all of the fields
and their derivatives and the integral runs over some spacetime manifold
$M$ \cite{Dyer2009}. When we vary $S\left[\phi\right]$ to obtain
the equations of motion, we want to be left with an expression of
the form 
\begin{equation}
\int_{M}d^{D}x\,\sqrt{-g}\,\left(\text{eom}_{i}\left[\phi\right]\delta\phi^{i}\right)
\end{equation}
which allows to deduce 
\begin{equation}
\text{eom}_{i}\left[\phi\right]=0\,\,\,\,\,\,\,\,\,\,\,\,\,\,\,\,\forall i.\label{eq:eoms}
\end{equation}
However, to isolate $\delta\phi^{i},$ it is in general necessary
to apply partial integration, which gives rise to surface terms evaluated
at $\partial M$: 
\begin{equation}
\delta S\left[\phi\right]=\int_{M}d^{D}x\,\sqrt{-g}\,\left(\text{eom}_{i}\left[\phi\right]\delta\phi^{i}\right)+\int_{\partial M}d^{D-1}x\,\sqrt{\mid h\mid}\,X\left(\left[\phi\right],\left[\delta\phi\right],x^{\mu}\right)
\end{equation}
Here, $h$ denotes the determinant of the induced metric on $\partial M$.
For the variational principle to be well-posed, we need (\ref{eq:eoms})
to be sufficient for making $\delta S\left[\phi\right]$ vanish, so
we need some additional strategy to eliminate $X\left(\left[\phi\right],\left[\delta\phi\right],x^{\mu}\right).$
There are two possibilities:
\begin{enumerate}
\item Impose boundary conditions at $\partial M$ for the fields $\phi^{i}$
and/or their derivatives.
\item Add an explicit boundary term of the form $S_{B}\left[\phi\right]=\int_{\partial M}d^{D-1}x\,\sqrt{\mid h\mid}\,\mathcal{L}_{B}\left(\left[\phi\right],x^{\mu}\right)$
that cancels $X$, i. e. 
\begin{equation}
\delta S_{B}\left[\phi\right]=-\int_{\partial M}d^{D-1}x\,\sqrt{\mid h\mid}\,X\left(\left[\phi\right],\left[\delta\phi\right],x^{\mu}\right)
\end{equation}
\end{enumerate}
It is instructive to analyze those two strategies with the help of
the simple example of a free real scalar field in flat space: We start
with the action 
\begin{equation}
S\left[\phi\right]=\int_{M}d^{4}x\,\,\frac{1}{2}\phi\boxempty\phi.
\end{equation}
Varying yields straightforwardly 
\begin{align}
\delta S\left[\phi\right] & =\int_{M}d^{4}x\,\,\frac{1}{2}\left(\delta\phi\boxempty\phi+\phi\boxempty\delta\phi\right)=\int_{M}d^{4}x\,\,\frac{1}{2}\left(\delta\phi\boxempty\phi-\partial_{\mu}\phi\partial^{\mu}\delta\phi\right)+\int_{\partial M}d^{D-1}x\,\,\frac{1}{2}n_{\mu}\phi\partial^{\mu}\delta\phi\nonumber \\
 & =\int_{M}d^{4}x\,\,\delta\phi\boxempty\phi+\int_{\partial M}d^{D-1}x\,\,\frac{1}{2}n_{\mu}\left(\phi\partial^{\mu}\delta\phi-\partial^{\mu}\phi\,\delta\phi\right),
\end{align}
where $n_{\mu}$ denotes the normal vector to the boundary. We can
read off directly 
\begin{equation}
\text{eom}\left[\phi\right]=\boxempty\phi,\,\,\,\,\,\,\,\,\,\,\,\,\,\,X\left(\left[\phi\right],\left[\delta\phi\right]\right)=\frac{1}{2}n_{\mu}\left(\phi\partial^{\mu}\delta\phi-\partial^{\mu}\phi\,\delta\phi\right)
\end{equation}
One way to proceed now would be to apply strategy one and impose Neumann
boundary conditions for $\phi$, i. e. demand 
\begin{equation}
n_{\mu}\partial^{\mu}\phi\mid_{\partial M}=0.
\end{equation}
This would imply $n_{\mu}\partial^{\mu}\delta\phi\mid_{\partial M}=0$
as well and hence kill the full boundary term. However, Neumann boundary
conditions allow only for a subspace of the physically interesting
solutions, we might be also interested in field configurations which
obey the Dirichlet condition $\delta\phi\mid_{\partial M}=0$. But
this does not kill the full boundary term, we are left with
\begin{equation}
X_{rem}\left(\left[\phi\right],\left[\delta\phi\right]\right)=\frac{1}{2}n_{\mu}\phi\partial^{\mu}\delta\phi.
\end{equation}

Let us examine the possibilities of strategy two: We would need to
be able to write $X\left(\left[\phi\right],\left[\delta\phi\right]\right)$
as $\delta$ of something, or in other words, if we regard the space
of fields $\phi\left(x^{\mu}\right)$ as a Banach manifold with $\delta$
being the exterior derivative, then $X$ must be an exact one-form.\footnote{In the case of string field theory, this is subtle because we do not
have a precise and consistent definition of the Hilbert space of string
fields available yet. With a slight abuse of language, we will still
use the terminology of differential forms for expressions using the
variational $\delta$. } A necessary condition for that to be true is that $X$ is closed,
$\delta X=0$. By direct calculation we find that 
\begin{equation}
\delta X\left(\left[\phi\right],\left[\delta\phi\right]\right)=\frac{1}{2}n_{\mu}\left(\delta\phi\land\partial^{\mu}\delta\phi-\partial^{\mu}\delta\phi\land\delta\phi\right)=\delta\phi\land n_{\mu}\partial^{\mu}\delta\phi\neq0,
\end{equation}
so strategy two alone is not sufficient to kill the full boundary
term. However, after imposing Dirichlet boundary conditions, we find
\begin{equation}
X_{rem}\left(\left[\phi\right],\left[\delta\phi\right]\right)=\frac{1}{2}n_{\mu}\phi\partial^{\mu}\delta\phi=\frac{1}{2}n_{\mu}\partial^{\mu}\left(\phi\delta\phi\right)=\delta\left(\frac{1}{4}n_{\mu}\partial^{\mu}\phi^{2}\right),
\end{equation}
so adding the term 
\begin{equation}
S_{B}\left[\phi\right]=-\frac{1}{4}\int_{\partial M}d^{3}x\,n_{\mu}\partial^{\mu}\phi^{2}
\end{equation}
to the action eliminates all boundary contributions. To summarize,
there are two requirements that a ``good'' set of boundary conditions
should satisfy: It should be general enough to include all kinds of
physically relevant solutions, but at the same time strict enough
to make $X_{rem}$ $\delta$-exact. 

\section{Kinetic boundary term}

We want to apply all of that now to the case of open string field
theory and focus first of all on the kinetic term
\begin{equation}
S_{kin}\left(\Psi\right)=-\frac{1}{2}\Braket{\Psi,Q\Psi}.
\end{equation}
To start, we need to make the dependence on the spacetime coordinates
and derivatives explicit, i. e. pass from momentum space to position
space. A general Fock space state has the form 
\begin{equation}
\Psi=\alpha_{-n_{1}}^{\mu_{1}}...\alpha_{-n_{r}}^{\mu_{r}}b_{-m_{1}}...b_{-m_{s}}c_{-p_{1}}...c_{-p_{t}}\Ket{0,k},\,\,\,\,\,\,\,\,\,n_{i}\geq1,\,\,\,\,m_{i}\geq2,\,\,\,\,p_{i}\geq-1\label{eq:Fock state}
\end{equation}
while its BPZ-conjugate is given by 
\begin{equation}
\Psi^{*}=\left(-1\right)^{\Sigma n_{i}+\Sigma m_{j}+\Sigma p_{k}+r+t}\Bra{0,k}\alpha_{n_{1}}^{\mu_{1}}...\alpha_{n_{r}}^{\mu_{r}}b_{m_{1}}...b_{m_{s}}c_{p_{1}}...c_{p_{t}}
\end{equation}
The vacua form the basic overlap 
\begin{equation}
\Bra{0,k}c_{-1}c_{0}c_{1}\Ket{0,k'}=\left(2\pi\right)^{26}\delta^{\left(26\right)}\left(k+k'\right).\label{eq:basic overlap}
\end{equation}
with the momentum dependent vacuum defined as 
\begin{equation}
\Ket{0,k}=:e^{ik_{\mu}\hat{X}^{\mu}\left(0,0\right)}:\Ket{0}=e^{ik_{\mu}\hat{x}^{\mu}}\Ket{0},
\end{equation}
($\Ket{0}$ denotes the $SL\left(2\right)$-invariant vacuum). The
zero-mode factor of our Hilbert space is just $L^{2}\left(M\right)$,
where $M$ is the 26-dimensional spacetime, so the operator $\hat{x}^{\mu}$
can act by multiplication with $x^{\mu}$ (We assume that $M$ is
just flat space with no background fields). The overlap (\ref{eq:basic overlap})
can now be represented as an integral over spacetime:
\begin{align}
\Bra{0,k}c_{-1}c_{0}c_{1}\Ket{0,k'} & =\int d^{26}x\,\Bra{0}e^{ik_{\mu}x^{\mu}}c_{-1}c_{0}c_{1}e^{ik'_{\nu}x{}^{\nu}}\Ket{0}'\nonumber \\
 & =\int d^{26}x\,e^{i\left(k_{\mu}+k'_{\mu}\right)x^{\mu}}\Bra{0}c_{-1}c_{0}c_{1}\Ket{0}'=\left(2\pi\right)^{26}\delta^{\left(26\right)}\left(k+k'\right)
\end{align}
Here, $\Bra{0}\Ket{0}'$ denotes the reduced BPZ-product without the
$x$-integration, which is written out explicitly. The position dependent
form of the string field is a linear superposition of Fock space states
(\ref{eq:Fock state}) of the form 
\begin{align}
\Ket{\Psi} & =\int\frac{d^{26}k}{\left(2\pi\right)^{26}}f\left(k\right)\alpha_{-n_{1}}^{\mu_{1}}...\alpha_{-n_{r}}^{\mu_{r}}b_{-m_{1}}...b_{-m_{s}}c_{-p_{1}}...c_{-p_{t}}\Ket{0,k}\nonumber \\
 & =\alpha_{-n_{1}}^{\mu_{1}}...\alpha_{-n_{r}}^{\mu_{r}}b_{-m_{1}}...b_{-m_{s}}c_{-p_{1}}...c_{-p_{t}}F\left(x\right)\Ket{0}\equiv\Ket{\Psi\left(x\right)}
\end{align}
where $F\left(x\right)$ is the Fourier transform of $f\left(k\right)$.\footnote{In the presence of a boundary, the integral over $k$ might get replaced
by a discrete sum.} The zero mode $p_{\mu}$ can now be represented as a derivative:
\begin{equation}
\alpha_{0\mu}=\sqrt{2\alpha'}p_{\mu}=-i\sqrt{2\alpha'}\partial_{\mu}.
\end{equation}
This is the only place where a spacetime derivative is appearing,
so from on now on we can focus solely on terms involving $\alpha_{0\mu}$. 

The relation which will be modified is the cyclicity condition of
$Q$\footnote{We assume that $\Psi$ has ghost number one such that Grassmann signs
can be omitted.},
\begin{equation}
\Braket{\Psi_{1},Q\Psi_{2}}=\Braket{Q\Psi_{1},\Psi_{2}}.
\end{equation}
From 
\begin{equation}
Q=\sum_{n}c_{-n}L_{n}^{\left(m\right)}+\text{ghosts}
\end{equation}
we can infer that there will be terms with one derivative proportional
to $-i\sqrt{2\alpha'}\sum_{n\neq0}c_{-n}\alpha_{n}^{\mu}\partial_{\mu}$
and one term with two derivatives given by $-\alpha'c_{0}\partial^{\mu}\partial_{\mu}$.
Examining first the two-derivative term, we see that it can be commuted
through until it hits the $x$-dependent function; explicitly we get
\begin{align}
-\alpha'\int_{M}d^{26}x\,\Bra{\Psi_{1}\left(x\right)}c_{0}\partial_{\mu}\partial^{\mu}\Ket{\Psi_{2}\left(x\right)}'= & -\alpha'\int_{M}d^{26}x\,\partial_{\mu}\left(\Bra{\Psi_{1}\left(x\right)}c_{0}\Ket{\partial^{\mu}\Psi_{2}\left(x\right)}'\right)\,\nonumber \\
 & +\alpha'\int_{M}d^{26}x\,\Bra{\partial_{\mu}\Psi_{1}\left(x\right)}c_{0}\Ket{\partial^{\mu}\Psi_{2}\left(x\right)}'\nonumber \\
=-\alpha'\int_{\partial M}d^{25}x\,n_{\mu}\left(\Bra{\Psi_{1}\left(x\right)}c_{0}\Ket{\partial^{\mu}\Psi_{2}\left(x\right)}'\right) & +\alpha'\int_{M}d^{26}x\,\partial^{\mu}\left(\Bra{\partial_{\mu}\Psi_{1}\left(x\right)}c_{0}\Ket{\Psi_{2}\left(x\right)}'\right)\nonumber \\
 & -\alpha'\int_{M}d^{26}x\,\Bra{\partial^{\mu}\partial_{\mu}\Psi_{1}\left(x\right)}c_{0}\Ket{\Psi_{2}\left(x\right)}'\nonumber \\
=-\alpha'\int_{\partial M}d^{25}x\,n_{\mu}\left(\Bra{\Psi_{1}\left(x\right)}c_{0}\Ket{\partial^{\mu}\Psi_{2}\left(x\right)}'\right) & +\alpha'\int_{\partial M}d^{25}x\,n_{\mu}\left(\Bra{\partial^{\mu}\Psi_{1}\left(x\right)}c_{0}\Ket{\Psi_{2}\left(x\right)}'\right)\nonumber \\
 & -\alpha'\int_{M}d^{26}x\,\Bra{\partial^{\mu}\partial_{\mu}\Psi_{1}\left(x\right)}c_{0}\Ket{\Psi_{2}\left(x\right)}'
\end{align}
and find two boundary terms in analogy with section two. The one-derivative
terms yield 
\begin{align}
-i\sqrt{2\alpha'}\sum_{n\neq0}\int_{M}d^{26}x\,\Bra{\Psi_{1}\left(x\right)}c_{-n}\alpha_{n}^{\mu}\partial_{\mu}\Ket{\Psi_{2}\left(x\right)}'\nonumber \\
=-i\sqrt{2\alpha'}\sum_{n\neq0}\int_{\partial M}d^{25}x\,n_{\mu}\Bra{\Psi_{1}\left(x\right)}c_{-n}\alpha_{n}^{\mu}\Ket{\Psi_{2}\left(x\right)}' & +i\sqrt{2\alpha'}\sum_{n\neq0}\int_{M}d^{26}x\,\Bra{\partial_{\mu}\Psi_{1}\left(x\right)}c_{-n}\alpha_{n}^{\mu}\Ket{\Psi_{2}\left(x\right)}'.
\end{align}
Substituting $\Psi_{1}=\Psi,$ $\Psi_{2}=\delta\Psi$ all boundary
terms together read 
\begin{equation}
\int_{\partial M}d^{25}x\,n_{\mu}\left(\alpha'\Bra{\partial^{\mu}\Psi\left(x\right)}c_{0}\Ket{\delta\Psi\left(x\right)}'-\alpha'\Bra{\Psi\left(x\right)}c_{0}\Ket{\partial^{\mu}\delta\Psi\left(x\right)}'-i\sqrt{2\alpha'}\sum_{n\neq0}\Bra{\Psi\left(x\right)}c_{-n}\alpha_{n}^{\mu}\Ket{\delta\Psi\left(x\right)}'\right).\label{eq:naive boundary term}
\end{equation}
If we proceed as in the example and apply Dirichlet (Neumann) boundary
conditions, the first and last (first and second, respectively) term
vanish and the remainder is an exact one-form:
\begin{equation}
-\int_{\partial M}d^{25}x\,n_{\mu}\left(\alpha'\Bra{\Psi\left(x\right)}c_{0}\Ket{\partial^{\mu}\delta\Psi\left(x\right)}'+i\sqrt{2\alpha'}\sum_{n\neq0}\Bra{\Psi\left(x\right)}c_{-n}\alpha_{n}^{\mu}\Ket{\delta\Psi\left(x\right)}'\right).
\end{equation}
It can be canceled by adding the boundary term 
\begin{align}
S_{B} & =\int_{\partial M}d^{25}x\,n_{\mu}\left(\alpha'\Bra{\Psi\left(x\right)}c_{0}\Ket{\partial^{\mu}\Psi\left(x\right)}'+i\sqrt{\frac{\alpha'}{2}}\sum_{n\neq0}\Bra{\Psi\left(x\right)}c_{-n}\alpha_{n}^{\mu}\Ket{\Psi\left(x\right)}'\right)\nonumber \\
 & =-\int_{\partial M}d^{25}x\,n_{\mu}\Bra{\Psi\left(x\right)}\left(c\partial X^{\mu}\right)_{0}\Ket{\Psi\left(x\right)}'\label{eq:GH-like term}
\end{align}
to the action. This Gibbons-Hawking-type term for the kinetic part
is one of the main results. 

\section{Breaking of conformal invariance\label{sec:Breaking-of-conformal}}

The derivation in the last section is probably the simplest and fastest
way to compute the boundary term, however, it does not connect naturally
to the operator formalism of the CFT. Here we will present an alternative
method based on a path integral argument. 

We have seen that the cyclicity relation of $Q,$
\begin{equation}
\Braket{\Psi_{1},Q\Psi_{2}}=\Braket{Q\Psi_{1},\Psi_{2}},\label{eq:cyclicity of Q}
\end{equation}
is violated in the presence of a boundary in target space. In the
CFT formalism this relation can be derived by writing 
\begin{equation}
\Braket{\Psi_{1},Q\Psi_{2}}=\Bra{0}\Psi_{1}^{*}\left(\infty\right)\oint_{C\left(0\right)}\frac{dz}{2\pi i}j_{B}\left(z\right)\Psi_{2}\left(0\right)\Ket{0}
\end{equation}
and deforming the integration contour until it encycles the other
insertion at infinity.\footnote{Here and in the whole rest of the paper the doubling trick is used.}
Since contour deformation is a mathematically well-established technique
valid in any two dimensional CFT, this is in sharp contradiction with
our result from the last section. 

The contradiction is resolved in \cite{Kraus2002}, where the authors
show that conformal invariance is broken by the target space boundary.
In a nutshell, their argument goes as follows: By using the explicit
expression of the worldsheet energy momentum tensor\footnote{Here, the worldsheet is a flat strip with coordinates $\sigma$, $\tau$
and a flat metric $\eta{}_{ab}$. For later purposes we assume that
in addition to $0\leq\sigma\leq\pi$ we have $\tau_{min}\leq\tau\leq\tau_{max}$,
such that the worldsheet has a finite area $V_{W}=\pi\left(\tau_{max}-\tau_{min}\right).$
In the end of the calculation we can savely take the limit $\tau_{max/min}\rightarrow\pm\infty$.} 
\begin{equation}
T_{ab}=-\frac{1}{\alpha'}\left(\partial_{a}X^{\mu}\partial_{b}X_{\mu}-\frac{1}{2}\eta_{ab}\eta^{cd}\partial_{c}X^{\mu}\partial_{d}X_{\mu}\right)
\end{equation}
we can write
\begin{equation}
\partial^{a}T_{ab}=-\frac{1}{\alpha'}\partial^{a}\partial_{a}X^{\mu}\partial_{b}X_{\mu}=-2\pi\frac{\delta S_{P}}{\delta X^{\mu}}\partial_{b}X^{\mu}\label{eq:conservation of T_ab}
\end{equation}
where $S_{P}$ denotes the Polyakov action. We see that $T_{ab}$
is classically conserved on-shell, as expected. To make sense of the
r. h. s. in the quantum theory, it first has to be normal-ordered,
i. e. using the standard prescription
\begin{equation}
:X^{\mu}\left(\sigma\right)X^{\nu}\left(\sigma'\right):=X^{\mu}\left(\sigma\right)X^{\nu}\left(\sigma'\right)+\frac{\alpha'}{2}\eta^{\mu\nu}\text{ln}\left(\mid\sigma-\sigma'\mid^{2}\right)
\end{equation}
we get 
\begin{equation}
:\frac{\delta S_{P}}{\delta X_{\mu}}\partial_{b}X_{\mu}:=\frac{\delta S_{P}}{\delta X^{\mu}}\partial_{b}X^{\mu}-26\partial_{b}\delta^{\left(2\right)}\left(0\right).
\end{equation}
Now, Eq. (\ref{eq:conservation of T_ab}) can be written as an operator
equation in the path integral as 
\begin{align}
\Braket{:\partial^{a}T_{ab}\left(\sigma\right):\mathcal{O}\left(\sigma_{1}\right)...\mathcal{O}\left(\sigma_{n}\right)} & =-2\pi\int\mathcal{D}X\,:\frac{\delta S_{P}}{\delta X^{\mu}}\partial_{b}X^{\mu}:\left(\sigma\right)\mathcal{O}\left(\sigma_{1}\right)...\mathcal{O}\left(\sigma_{n}\right)e^{-S_{P}}\nonumber \\
 & =2\pi\int\mathcal{D}X\,\frac{\delta}{\delta X^{\mu}\left(\sigma\right)}\left(\partial_{b}X^{\mu}\left(\sigma\right)\mathcal{O}\left(\sigma_{1}\right)...\mathcal{O}\left(\sigma_{n}\right)e^{-S_{P}}\right).\label{eq:path integral}
\end{align}
We assumed for simplicity that there are no contact terms, $\sigma\neq\sigma_{i}$.
In ordinary cases one would now argue that the path integral of the
total functional derivative vanishes, but there may be contributions
from the boundary of spacetime: If we expand $X^{\mu}$ in mutually
orthogonal modes with $V_{W}$ being the volume of the worldsheet,
\begin{equation}
X^{\mu}=x^{\mu}+\sum_{n\neq0}x_{n}^{\mu}X_{n}\left(\sigma\right),\,\,\,\,\,\,\,\,\,\int d^{2}\sigma\,X_{n}X_{m}=\delta_{mn},\,\,\,\,\,\,\,\,\,\,\frac{\delta}{\delta X^{\mu}\left(\sigma\right)}=V_{W}^{-1}\frac{\partial}{\partial x^{\mu}}+\sum_{n\neq0}X_{n}\left(\sigma\right)\frac{\partial}{\partial x_{n}^{\mu}}
\end{equation}
then the boundary in field configuration space corresponds to large
values of the coefficients $x^{\mu}$, $x_{n}^{\mu}$. The non-constant
modes $X_{n}$ are exponentially suppressed by $e^{-S_{P}}$, but
this does not hold for the constant mode since $X^{\mu}$ appears
only derivated in $S_{P}$. So we are only allowed to neglect the
non-constant modes in (\ref{eq:path integral}) and are left with
\begin{align}
\Braket{\partial^{a}T_{ab}\left(\sigma\right)\mathcal{O}\left(\sigma_{1}\right)...\mathcal{O}\left(\sigma_{n}\right)} & =2\pi V_{W}^{-1}\int_{M}d^{26}x\,\frac{\partial}{\partial x^{\mu}}\Braket{\partial_{b}X^{\mu}\left(\sigma\right)\mathcal{O}\left(\sigma_{1}\right)...\mathcal{O}\left(\sigma_{n}\right)}'\nonumber \\
 & =2\pi V_{W}^{-1}\int_{\partial M}d^{25}x\,n_{\mu}\Braket{\partial_{b}X^{\mu}\left(\sigma\right)\mathcal{O}\left(\sigma_{1}\right)...\mathcal{O}\left(\sigma_{n}\right)}'.\label{eq:general boundary formula}
\end{align}

This violation of the conformal Ward identity will give rise to extra
terms in the contour deformation: We will focus solely on the spacetime-dependent
part and ignore the ghost contribution to get
\begin{align}
\Bra{0}\Psi_{1}^{*}\left(\infty\right)\left(\oint_{C_{1}}\frac{dz}{2\pi i}\left(cT^{\left(m\right)}\right)-\oint_{C_{2}}\frac{d\overline{z}}{2\pi i}\left(\overline{c}\overline{T}^{\left(m\right)}\right)\right)\left(z,\overline{z}\right)\Psi_{2}\left(0\right)\Ket{0}\nonumber \\
\subset\Bra{0}\Psi_{1}^{*}\left(\infty\right)\oint_{C\left(0\right)}\frac{dz}{2\pi i}j_{B}\left(z\right)\Psi_{2}\left(0\right)\Ket{0}
\end{align}
where we made the doubling trick explicit (Here, $C_{1/2}$ denotes
the upper/lower semicircle of the contour.). Now we can use the complex
divergence theorem 
\begin{equation}
\int_{R}d^{2}z\,\left(\partial v^{z}+\overline{\partial}v^{\overline{z}}\right)=i\oint_{\partial R}\left(v^{z}d\overline{z}-v^{\overline{z}}dz\right)\label{eq:doubling trick}
\end{equation}
to rewrite the integral: Setting $v^{z}=:$$\left(\overline{c}\overline{T}^{\left(m\right)}\right)$
supported in the lower and $v^{\overline{z}}=:\left(cT^{\left(m\right)}\right)$
supported in the upper half plane yields 
\begin{align}
 & \Bra{0}\Psi_{1}^{*}\left(\infty\right)\left(\oint_{C_{1}}\frac{dz}{2\pi i}\left(cT^{\left(m\right)}\right)-\oint_{C_{2}}\frac{d\overline{z}}{2\pi i}\left(\overline{c}\overline{T}^{\left(m\right)}\right)\right)\left(z,\overline{z}\right)\Psi_{2}\left(0\right)\Ket{0}\nonumber \\
= & \,\frac{1}{2\pi}\Bra{0}\Psi_{1}^{*}\left(\infty\right)\int_{Int\left(C\right)}d^{2}z\,\left(\overline{\partial}\left(cT^{\left(m\right)}\right)+\partial\left(\overline{c}\overline{T}^{\left(m\right)}\right)\right)\left(z,\overline{z}\right)\Psi_{2}\left(0\right)\Ket{0}\nonumber \\
= & \,\frac{1}{\pi}\Bra{0}\Psi_{1}^{*}\left(\infty\right)\int_{UpInt\left(C\right)}d^{2}z\,\overline{\partial}\left(cT^{\left(m\right)}\right)\left(z,\overline{z}\right)\Psi_{2}\left(0\right)\Ket{0}.\label{eq:T contour deformation}
\end{align}
In the last step we observed that both terms actually yield the same
integral, so we can restrict to the first term intergrated in the
upper half plane only. $c\left(z\right)$ is still holomorphic so
$\overline{\partial}c=0$, whileas $\overline{\partial}T^{\left(m\right)}$
will give rise to a non-trivial boundary contribution. For the cyclicity
relation (\ref{eq:cyclicity of Q}) we need to deform the contour
from an infinitesimal circle around zero to an infinitesimal circle
around infinity hence we have to cover the whole complex plane. In
the end we get 
\begin{equation}
\Braket{\Psi_{1},Q\Psi_{2}}-\Braket{Q\Psi_{1},\Psi_{2}}=\frac{1}{\pi}\Bra{0}\Psi_{1}^{*}\left(\infty\right)\int_{UHP}d^{2}z\,\overline{\partial}\left(cT^{\left(m\right)}\right)\left(z,\overline{z}\right)\Psi_{2}\left(0\right)\Ket{0}.
\end{equation}
To apply (\ref{eq:general boundary formula}) we need to transform
back to the sigma model coordinates: First, denoting $x=\text{Re }z$,
$y=\text{Im }z$ and using the fact that the energy-momentum-tensor
is still symmetric and traceless, we get 
\begin{equation}
\left(c\overline{\partial}T^{\left(m\right)}\right)\left(z,\overline{z}\right)=\frac{1}{4}\left(c^{x}+ic^{y}\right)\left(\partial^{j}T_{jx}-i\partial^{j}T_{jy}\right)\left(x,y\right).
\end{equation}
Now using the usual map from the strip to the upper half plane including
the Wick rotation $z=e^{i\left(-\tau+\sigma\right)}$, the expression
can be straightforwardly transformed into 
\begin{equation}
\left(c\overline{\partial}T^{\left(m\right)}\right)\left(z,\overline{z}\right)=\frac{1}{4}\left(c^{\sigma}-c^{\tau}\right)\left(\partial^{a}T_{a\sigma}-\partial^{a}T_{a\tau}\right)\left(\sigma,\tau\right).
\end{equation}
Finally we can insert into (\ref{eq:general boundary formula}), which
yields 
\begin{align}
 & \Braket{\Psi_{1},Q\Psi_{2}}-\Braket{Q\Psi_{1},\Psi_{2}}\nonumber \\
= & \,V_{W}^{-1}\int_{\partial M}d^{25}x\,n_{\mu}\Bra{0}\Psi_{1}^{*}\left(\tau_{max}\right)\int d^{2}\sigma\,\left(c^{\sigma}-c^{\tau}\right)\left(\partial_{\sigma}X^{\mu}-\partial_{\tau}X^{\mu}\right)\left(\sigma,\tau\right)\Psi_{2}\left(\tau_{min}\right)\Ket{0}'
\end{align}
where $d^{2}z=2d^{2}\sigma$ was used. The integral can now be computed
explicitly using the mode expansions 
\begin{equation}
\left(c^{\sigma}-c^{\tau}\right)\left(\sigma,\tau\right)\equiv-c^{-}\left(\sigma^{-}\right)=-\sum_{n}c_{n}e^{\left(n-1\right)i\sigma^{-}}
\end{equation}
\begin{equation}
\left(\partial_{\sigma}X^{\mu}-\partial_{\tau}X^{\mu}\right)\left(\sigma,\tau\right)\equiv-2\left(\partial_{-}X^{\mu}\right)\left(\sigma^{-}\right)=2\sum_{n\neq0}i\sqrt{\frac{\alpha'}{2}}\alpha_{n}^{\mu}e^{\left(n+1\right)i\sigma^{-}}+2i\alpha'p^{\mu}e^{i\sigma^{-}}.
\end{equation}
At this point it is advantageous to keep the zero mode explicit, as
we will see in a moment. We get 
\begin{equation}
\Braket{\Psi_{1},Q\Psi_{2}}-\Braket{Q\Psi_{1},\Psi_{2}}=\int_{\partial M}d^{25}x\,n_{\mu}\Bra{\Psi_{1}\left(x\right)}\left(-i\sqrt{2\alpha'}\sum_{n\neq0}c_{-n}\alpha_{n}^{\mu}-2i\alpha'c_{0}p^{\mu}\right)\Ket{\Psi_{2}\left(x\right)}'.
\end{equation}
This is already almost what we would expect from (\ref{eq:naive boundary term})
but one has to be careful how $p^{\mu}\propto\partial^{\mu}$ is acting.
The position of the insertion of $\left(c\partial X^{\mu}\right)_{0}$
was basically arbitrary since we could have as well started the calculation
from the other side, i. e. $Q$ acting on $\Psi_{1}$. Naively the
result would be the same because of momentum conservation, but we
have a reduced correlation function, where the direction normal to
the boundary is not integrated over. It hence makes a difference if
$\partial^{\mu}$ is acting on $\Psi_{1}$ or $\Psi_{2}$. 

The correct result can actually be derived via the same method: The
operator $p^{\mu}$ can be represented as a contour integral 
\begin{equation}
p^{\mu}\Psi_{2}\left(0\right)=i\frac{1}{\alpha'}\oint_{C\left(0\right)}\frac{dz}{2\pi i}\partial X^{\mu}\left(z\right)\Psi_{2}\left(0\right)
\end{equation}
where the doubling trick is used again. If we want $p^{\mu}$ act
on $\Psi_{1}\left(\infty\right)$, the contour has to be deformed
around the whole Riemann sphere and the difference can be expressed
with the help of (\ref{eq:doubling trick}) as 
\begin{align}
 & -2i\alpha'\int_{\partial M}d^{25}x\,n_{\mu}\left(\Bra{\Psi_{1}\left(x\right)}c_{0}p^{\mu}\Ket{\Psi_{2}\left(x\right)}'-p^{\mu}\Bra{\Psi_{1}\left(x\right)}c_{0}\Ket{\Psi_{2}\left(x\right)}'\right)\nonumber \\
=\, & \frac{1}{\pi}\int_{\partial M}d^{25}x\,n_{\mu}\int d^{2}z\,\Bra{\Psi_{1}\left(x\right)}c_{0}\overline{\partial}\partial X^{\mu}\left(z,\overline{z}\right)\Ket{\Psi_{2}\left(x\right)}',\label{eq:intermediate step}
\end{align}
using the same arguments as in (\ref{eq:T contour deformation}).
This expression would normally vanish as it contains the classical
equations of motion inside a correlation function. In the presence
of a boundary however, we find analogously to (\ref{eq:path integral})
\begin{align}
\Braket{\partial_{+}\partial_{-}X^{\mu}\left(\sigma\right)\mathcal{O}\left(\sigma_{1}\right)...\mathcal{O}\left(\sigma_{n}\right)} & =\frac{\pi}{2}\alpha'\int\mathcal{D}X\,\frac{\delta S_{P}}{\delta X^{\mu}}\left(\sigma\right)\mathcal{O}\left(\sigma_{1}\right)...\mathcal{O}\left(\sigma_{n}\right)e^{-S_{P}}\nonumber \\
 & =-\frac{\pi}{2}\alpha'\int\mathcal{D}X\,\frac{\delta}{\delta X^{\mu}\left(\sigma\right)}\left(\mathcal{O}\left(\sigma_{1}\right)...\mathcal{O}\left(\sigma_{n}\right)e^{-S_{P}}\right)\nonumber \\
 & =-\frac{\pi}{2}\alpha'V_{W}^{-1}\int_{\partial M}d^{25}x\,n_{\mu}\Braket{\mathcal{O}\left(\sigma_{1}\right)...\mathcal{O}\left(\sigma_{n}\right)}'.
\end{align}
If we write (\ref{eq:intermediate step}) again as a 26-dimensional
integral to have a full correlation function we can insert and get
\begin{align}
 & \frac{1}{\pi}\int_{M}d^{26}x\,\partial_{\mu}\int d^{2}z\,\Bra{\Psi_{1}\left(x\right)}c_{0}\overline{\partial}\partial X^{\mu}\left(z,\overline{z}\right)\Ket{\Psi_{2}\left(x\right)}'\nonumber \\
= & \,\frac{1}{\pi}\partial_{\mu}\int d^{2}z\,\Bra{\Psi_{1}\left(x\right)}c_{0}\overline{\partial}\partial X^{\mu}\left(z,\overline{z}\right)\Ket{\Psi_{2}\left(x\right)}\nonumber \\
= & \,-\frac{\alpha'}{2}V_{W}^{-1}\int_{\partial M}d^{25}x\,n_{\mu}\partial^{\mu}\int d^{2}z\,\Bra{\Psi_{1}\left(x\right)}c_{0}\Ket{\Psi_{2}\left(x\right)}'\nonumber \\
= & \,-\alpha'\int_{\partial M}d^{25}x\,n_{\mu}\partial^{\mu}\left(\Bra{\Psi_{1}\left(x\right)}c_{0}\Ket{\Psi_{2}\left(x\right)}'\right).\label{eq:difference}
\end{align}
(In the last line we used $\int d^{2}z=2\int d^{2}\sigma=2V_{W}$.)
This in general non-zero, so the only chance to get a well-defined
result is that the operator $c\partial X^{\mu}$ shall act \emph{symmetrically
}on all of the string fields involved:
\begin{equation}
\int_{\partial M}d^{25}x\,n_{\mu}\left(\alpha'\partial^{\mu}\Bra{\Psi_{1}\left(x\right)}c_{0}\Ket{\Psi_{2}\left(x\right)}'-\alpha'\Bra{\Psi_{1}\left(x\right)}c_{0}\partial^{\mu}\Ket{\Psi_{2}\left(x\right)}'-i\sqrt{2\alpha'}\sum_{n\neq0}\Bra{\Psi_{1}\left(x\right)}c_{-n}\alpha_{n}^{\mu}\Ket{\Psi_{2}\left(x\right)}'\right).
\end{equation}
(The minus sign between the first two terms comes from commuting $c_{0}$
with $\Psi_{1}$.) We see that the difference between the first two
terms is just (\ref{eq:difference}), as expected. If we now set $\Psi_{1}=\Psi$
and $\Psi_{2}=\delta\Psi$ we precisely reproduce (\ref{eq:naive boundary term}). 

\section{Cubic boundary term\label{sec:Cubic-boundary-term}}

In this section we discuss the cubic vertex 
\begin{equation}
S_{int}\left(\Psi\right)=-\frac{1}{3}\Braket{\Psi,\Psi\ast\Psi}.\label{eq:original 3-vertex}
\end{equation}
and possible boundary contribution that we expect to get from it.
This is actually very subtle since the interaction is non-local, whileas
the boundary is localized in spacetime. Traditionally, the vertex
is defined as 
\begin{equation}
\Braket{\Psi_{1},\Psi_{2}\ast\Psi_{3}}=:\Braket{f_{1}\circ\Psi_{1}\left(0\right)f_{2}\circ\Psi_{2}\left(0\right)f_{3}\circ\Psi_{3}\left(0\right)}\label{eq:Three vertex}
\end{equation}
with 
\begin{equation}
f_{j}\left(z\right)=\text{tan}\left(\frac{2}{3}\left(\text{arctan}\,z-\pi+\frac{j\pi}{2}\right)\right).
\end{equation}
We will however see that this definition has to be modified in the
presence of a boundary. The cyclicity relation 
\begin{equation}
\Braket{f_{1}\circ\Psi_{1}\left(0\right)f_{2}\circ\Psi_{2}\left(0\right)f_{3}\circ\Psi_{3}\left(0\right)}=\Braket{f_{1}\circ\Psi_{3}\left(0\right)f_{2}\circ\Psi_{1}\left(0\right)f_{3}\circ\Psi_{2}\left(0\right)}
\end{equation}
is proven using the map 
\begin{equation}
\tilde{f}\left(z\right)=\text{tan}\left(\text{arctan}\,z+\frac{\pi}{3}\right)=\frac{z+\sqrt{3}}{1-\sqrt{3}z}
\end{equation}
which fulfills $\tilde{f}\circ f_{j}\left(z\right)=f_{j+1}\left(z\right)$
modulo three. This transformation is not expected to give rise to
any boundary term which can be argued as follows\footnote{GS wants to thank Barton Zwiebach for this observation.}:
Consider the unit disc coordinate 
\begin{equation}
\xi=\frac{1+iz}{1-iz}
\end{equation}
where the functions defining the three-vertex take the form 
\begin{equation}
g_{1}\left(\xi\right)=e^{-\frac{2\pi i}{3}}\xi^{\frac{2}{3}},\,\,\,\,\,\,\,\,\,\,\,g_{2}\left(\xi\right)=\xi^{\frac{2}{3}},\,\,\,\,\,\,\,\,\,\,\,\,g_{3}\left(\xi\right)=e^{\frac{2\pi i}{3}}\xi^{\frac{2}{3}}.
\end{equation}
Here, cyclicity is shown via a simple rotation by $\frac{2\pi}{3}$
and rotational symmetry should not be broken by the target space boundary. 

To analyze where boundary terms may arise, let us consider an arbitrary
tachyon field $t\left(x\right)=\int\frac{d^{26}k}{\left(2\pi\right)^{26}}T\left(k\right)ce^{ik^{\mu}x_{\mu}}\Ket{0}$:
One gets 
\begin{equation}
S_{int}\left(t\left(x\right)\right)\propto\int d^{26}x\,\left(e^{-\alpha'\text{ln}\frac{4}{3\sqrt{3}}\square}t\left(x\right)\right)^{3}\label{eq:tachyon vertex}
\end{equation}
(for a derivation see for instance \cite{Ohmori2001}). It is instructive
to expand this expression in powers of $\alpha'$ and check different
orders separately. While the first order is very similar to the example
in section \ref{sec:General-philosophy}, at second order we get up
to prefactors
\begin{equation}
\int d^{26}x\,\left(\frac{1}{2}t^{2}\left(x\right)\Square^{2}t\left(x\right)+t\left(x\right)\left(\Square t\left(x\right)\right)^{2}\right).
\end{equation}
We can now straightforwardly compute the associated boundary term
as outlined in section \ref{sec:General-philosophy} and get 
\begin{equation}
X=\int_{\partial M}d^{25}x\,n_{\mu}\left(\frac{1}{2}t^{2}\partial^{\mu}\Square\delta t-t\partial^{\mu}t\Square\delta t+3t\Square t\partial^{\mu}\delta t+\partial_{\nu}t\partial^{\nu}t\partial^{\mu}\delta t-3t\partial^{\mu}\Square t\delta t-4\partial^{\mu}t\Square t\delta t\right),
\end{equation}
\begin{equation}
\delta X=\int_{\partial M}d^{25}x\,n_{\mu}\left(4t\delta t\wedge\partial^{\mu}\Square\delta t+4\partial^{\mu}t\delta t\wedge\Square\delta t+8\Square t\delta t\wedge\partial^{\mu}\delta t+4t\Square\delta t\wedge\partial^{\mu}\delta t+2\partial_{\nu}t\partial^{\nu}\delta t\land\partial^{\mu}\delta t\right).
\end{equation}
Even after applying boundary conditions, $\delta X$ is not zero:
For Neumann conditions we are left with the first term and Dirichlet
conditions would only eliminate the first three terms. The only possibility
we have is to apply either Neumann and Dirichlet conditions simultaneously
or put a condition on the third derivative of $\delta t$. Both are
physically problematic because important subsets of the solution space
are eliminated (see \cite{Moeller2002} for additional discussions).
Naively we would expect that at higher orders in $\alpha'$, higher
derivatives of $t\left(x\right)$ need to be constrained. At infinite
order, we would then end up with boundary conditions on all derivatives
of $t\left(x\right)$ and hence, in the case where the field $t\left(x\right)$
is analytic, it would be fully determined by the boundary conditions
alone. This would clearly be an unphysical situation and a reflection
of the above-mentioned fact that a sharply located boundary may not
be meaningful in a non-local theory. However, we must be careful when
summing up all infinitely many orders of $\alpha'$ because it might
change the behaviour: In \cite{Barnaby2008}, the authors consider
non-local kinetic operators in the context of the initial value problem,
where one would a priori expect an infinite number of possible initial
conditions. Indeed, it is shown that for a subclass of operators,
this is not the case and only a finite number of initial conditions
are sufficient. Although our problem is fundamentally different, it
is possible that upon including all orders, a similar simplification
occurs. Another possible route to follow is to consider an asymptotic
boundary in the spirit of conformal compactifications. Then boundary
conditions would be replaced by fall-off conditions on the string
field and the problem might be ameliorated. 

We see from this example that as we expected, boundary terms do not
arise from cyclic permutations; (\ref{eq:tachyon vertex}) is manifestly
cyclic even for three different tachyon fields. The crucial step is
to isolate the first string field, i. e. move the differential operator
$e^{-\alpha'\text{ln}\frac{4}{3\sqrt{3}}\square}$ to act on the other
two string fields. In fact, equation (\ref{eq:Three vertex}) should
already contain the boundary contribution: 
\begin{equation}
\Braket{\Psi_{1},\Psi_{2}\ast\Psi_{3}}=:\Braket{f_{1}\circ\Psi_{1}\left(0\right)f_{2}\circ\Psi_{2}\left(0\right)f_{3}\circ\Psi_{3}\left(0\right)}-X_{int}\left(\Psi_{1},\Psi_{2},\Psi_{3}\right)
\end{equation}
Indeed, if one defines 
\begin{equation}
S_{int}=:-\frac{1}{3}\Braket{f_{1}\circ\Psi\left(0\right)f_{2}\circ\Psi\left(0\right)f_{3}\circ\Psi\left(0\right)}
\end{equation}
as the cyclic expression, then varying yields 
\begin{equation}
\delta S_{int}=-\Braket{\delta\Psi,\Psi\ast\Psi}-X\left(\delta\Psi,\Psi,\Psi\right)
\end{equation}
as expected. 

Calculating $X$ explicitly for the full string field is troublesome,
from the above calculation we can expect that it will take a very
complicated form. One might be tempted to apply the function $f_{1}^{-1}\left(z\right)$
to (\ref{eq:Three vertex}), however, $f_{1}\left(z\right)$ does
not map the upper half plane to itself and is therefore not a symmetry
of the 3-vertex. One could use the sliver frame representation of
\cite{Schnabl:2005gv} where the vertex reads 
\begin{equation}
S_{int}=-\frac{1}{3}\Braket{s\circ\Psi\left(\frac{3\pi}{4}\right)s\circ\Psi\left(\frac{\pi}{4}\right)s\circ\Psi\left(-\frac{\pi}{4}\right)}_{C_{\pi}}.\label{eq:sliver frame 3-vertex}
\end{equation}
$s$ induces the scale transformation $w\rightarrow\frac{2}{3}w$
which is generated by the operator 
\begin{equation}
U_{3}=\left(\frac{2}{3}\right)^{\mathcal{L}_{0}}
\end{equation}
where $\mathcal{L}_{0}$ is the zero-mode of the energy momentum tensor
in the sliver frame. Writing (\ref{eq:sliver frame 3-vertex}) as
a two-point function leads to the expression 
\begin{equation}
S_{int}=-\frac{1}{3}\Braket{\Psi,U_{3}^{*}\left(s\circ\Psi\left(\frac{\pi}{4}\right)s\circ\Psi\left(-\frac{\pi}{4}\right)\right)}_{C_{\pi}}.
\end{equation}
This is exactly the step where the boundary terms come in, where $U_{3}$
is moved to the other side of the BPZ-product. To use the method of
section \ref{sec:Breaking-of-conformal} to calculate those boundary
terms, we would have to write $U_{3}$ as a contour integral. This
is not easy though, since $U_{3}$ is just the exponential of a contour
integral, hence we cannot proceed straightforwardly. 

To sum up, the cubic boundary terms exhibit serious difficulties,
on the one hand because of non-locality and on the other hand also
because of computational issues. A careful treatment is left for future
publications. 

\section{Conclusion and outlook}

In this paper we analyzed Witten's open string field theory in the
case where spacetime contains a boundary. We found non-trivial boundary
contributions from the variational principle both for the kinetic
as well as for the cubic term: In the former case, the cyclicity relation
for the BRST-operator is violated. We calculated a Gibbons-Hawking-like
surface term (\ref{eq:GH-like term}) that can be added to the action
to guarantee a well-defined variational principle if the string field
obeys either Neumann or Dirichlet conditions. The derivation was done
in two different ways, one rather heuristic way and one using the
breaking of conformal invariance in the path-integral formalism, both
yield the same result. For the cubic term we saw that boundary terms
arise when rewriting the manifestly cyclic form as a BPZ-product.
However, it is not straightforward to find suitable boundary or fall-off
conditions due to the non-locality of the three-vertex, which is one
of the most interesting prospects for further research. As another
future direction, it could be enlightening to extend the study to
more general backgrounds, i. e. linear dilaton or Liouville CFTs. 

\subsubsection*{\textcolor{black}{Acknowledgements}}

\textcolor{black}{GS wants to thank Martin Schnabl for suggesting
the topic and collaboration as well as Ted Erler, Atakan Firat, Carlo
Maccaferri, Riccardo Poletti, Alberto Ruffino, Jaroslav Scheinpflug,
Raphaela Wutte and Barton Zwiebach for useful discussions. This work
was co-funded by the European Union and supported by the Czech Ministry
of Education, Youth and Sports (Project No. FORTE -- CZ.02.01.01/00/22\_008/0004632).}

\subsubsection*{Note added}

Shortly after this work was completed, two other papers on the same
topic appeared which were created independently. In \cite{Firat2025}
the authors compute the kinetic boundary term for the closed string
and \cite{Maccaferri2025} deals with gauge invariance and boundary
modes. 

\bibliographystyle{plain}
\bibliography{String_Field_Theory}

\end{document}